# A General Expression for the Quintic Lovelock Tensor

C. C. Briggs
*Center for Academic Computing, Penn State University, University Park, PA 16802*

**Abstract.** A general expression is given for the quintic Lovelock tensor as well as for the coefficient of the quintic Lovelock Lagrangian in terms of the Riemann-Christoffel and Ricci curvature tensors and the Riemann curvature scalar for *n*-dimensional differentiable manifolds having a general linear connection.

This letter provides the interested reader with a general expression for the quintic Lovelock tensor $G_{(5)a}{}^{b}$ in terms of the Riemann-Christoffel curvature tensor[1]

$$R_{abc}{}^{d} \equiv 2\,(\partial_{[a}\,\Gamma_{b]}{}^{d}{}_{c} + \Gamma_{[a}{}^{d}{}_{|e|}\,\Gamma_{b]}{}^{e}{}_{c} + \Omega_{a}{}^{e}{}_{b}\,\Gamma_{e}{}^{d}{}_{c}), \qquad (1)$$

the Ricci curvature tensor

$$R_{a}{}^{b} \equiv R_{ca}{}^{bc} = -\,R_{ac}{}^{bc} = -\,R_{ca}{}^{cb} + 2\,(\nabla_{[c}\,Q_{a]}{}^{bc} + S_{ca}{}^{d}\,Q_{d}{}^{bc}), \qquad (2)$$

and the Riemann curvature scalar

$$R \equiv R_{a}{}^{a} = R_{ba}{}^{ab} = -\,R_{ab}{}^{ab} = -\,R_{ba}{}^{ba} \qquad (3)$$

using anholonomic coordinates for *n*-dimensional differentiable manifolds having a general linear connection, where $\partial_a$ is the Pfaffian derivative, $\Gamma_{a}{}^{b}{}_{c}$ the connection coefficient, $\Omega_{a}{}^{b}{}_{c}$ the object of anholonomity, $Q_{a}{}^{bc}$ the non-metricity tensor, and $S_{ab}{}^{c}$ the torsion tensor.

A general expression for the coefficient $L_{(5)}$ of the quintic Lovelock Lagrangian appears in the appendix.

Suchlike expressions for $G_{(5)a}{}^{b}$ and $L_{(5)}$ seem not to have appeared in the open literature hitherto, although isomers of the concomitants belonging to the $73^{rd}$ and $80^{th}$ terms of $L_{(5)}$ per Eq. (7) below have appeared in Zanon.[2]

In accordance with various general definitions given by Müller-Hoissen[3] and Verwimp,[4] the quintic Lovelock tensor $G_{(5)a}{}^{b}$ is given by the formula

$$G_{(5)a}{}^{b} = \frac{11!}{2^6 \times 5}\,\delta^{b}_{[a}\,R_{i_1 i_2}{}^{i_1 i_2}\,R_{i_3 i_4}{}^{i_3 i_4}\,R_{i_5 i_6}{}^{i_5 i_6}\,R_{i_7 i_8}{}^{i_7 i_8}\,R_{i_9 i_{10}]}{}^{i_9 i_{10}}, \qquad (4)$$

which comprises 39,916,800 unique covariant index permutations, of which but 596—together with numerical coefficients—suffice for rendering a general expression for $G_{(5)a}{}^{b}$, the final result (after substituting contractions and re-labeling indices) being given by

$$\begin{aligned}
G_{(5)a}{}^{b} = \tfrac{1}{10}\,(&- \delta^b_a R^5 + 40\,\delta^b_a R^3 R_d{}^c R_c{}^d - 10\,\delta^b_a R^3 R_{ef}{}^{cd} R_{cd}{}^{ef} - 160\,\delta^b_a R^2 R_d{}^c R_e{}^d R_c{}^e + 240\,\delta^b_a R^2 R_e{}^c R_f{}^d R_{cd}{}^{ef} + 240\,\delta^b_a R^2 R_d{}^c R_{fg}{}^{de} R_{ce}{}^{fg} + \qquad (5)\\
&+ 20\,\delta^b_a R^2 R_{ef}{}^{cd} R_{gh}{}^{ef} R_{cd}{}^{gh} - 80\,\delta^b_a R^2 R_{eg}{}^{cd} R_{ch}{}^{ef} R_{df}{}^{gh} - 240\,\delta^b_a R R_d{}^c R_c{}^d R_f{}^e R_e{}^f + 480\,\delta^b_a R R_d{}^c R_e{}^d R_f{}^e R_c{}^f - 1920\,\delta^b_a R R_d{}^c R_f{}^d R_g{}^e R_{ce}{}^{fg} +\\
&+ 120\,\delta^b_a R R_d{}^c R_c{}^d R_{gh}{}^{ef} R_{ef}{}^{gh} - 960\,\delta^b_a R R_d{}^c R_e{}^d R_{gh}{}^{ef} R_{cf}{}^{gh} - 480\,\delta^b_a R R_e{}^c R_f{}^d R_{gh}{}^{ef} R_{cd}{}^{gh} + 960\,\delta^b_a R R_e{}^c R_g{}^d R_{ch}{}^{ef} R_{df}{}^{gh} -\\
&- 960\,\delta^b_a R R_e{}^c R_g{}^d R_{dh}{}^{ef} R_{cf}{}^{gh} + 960\,\delta^b_a R R_d{}^c R_{cf}{}^{de} R_{hi}{}^{fg} R_{eg}{}^{hi} - 480\,\delta^b_a R R_d{}^c R_{fg}{}^{de} R_{hi}{}^{fg} R_{ce}{}^{hi} + 1920\,\delta^b_a R R_d{}^c R_{fh}{}^{de} R_{ci}{}^{fg} R_{eg}{}^{hi} -\\
&- 15\,\delta^b_a R R_{ef}{}^{cd} R_{cd}{}^{ef} R_{ij}{}^{gh} R_{gh}{}^{ij} + 240\,\delta^b_a R R_{ef}{}^{cd} R_{cg}{}^{ef} R_{ij}{}^{gh} R_{dh}{}^{ij} - 30\,\delta^b_a R R_{ef}{}^{cd} R_{gh}{}^{ef} R_{ij}{}^{gh} R_{cd}{}^{ij} + 480\,\delta^b_a R R_{ef}{}^{cd} R_{gi}{}^{ef} R_{cj}{}^{gh} R_{dh}{}^{ij} -\\
&- 240\,\delta^b_a R R_{eg}{}^{cd} R_{ci}{}^{ef} R_{dj}{}^{gh} R_{fh}{}^{ij} + 480\,\delta^b_a R R_{eg}{}^{cd} R_{ci}{}^{ef} R_{fj}{}^{gh} R_{dh}{}^{ij} + 640\,\delta^b_a R_d{}^c R_c{}^d R_f{}^e R_g{}^f R_e{}^g - 768\,\delta^b_a R_d{}^c R_e{}^d R_f{}^e R_g{}^f R_c{}^g -\\
&- 960\,\delta^b_a R_d{}^c R_c{}^d R_g{}^e R_h{}^f R_{ef}{}^{gh} + 3840\,\delta^b_a R_d{}^c R_e{}^d R_g{}^e R_h{}^f R_{cf}{}^{gh} + 1920\,\delta^b_a R_d{}^c R_g{}^d R_f{}^e R_h{}^f R_{ce}{}^{gh} - 960\,\delta^b_a R_d{}^c R_c{}^d R_f{}^e R_{hi}{}^{fg} R_{eg}{}^{hi} -\\
&- 160\,\delta^b_a R_d{}^c R_e{}^d R_c{}^e R_{hi}{}^{fg} R_{fg}{}^{hi} + 1920\,\delta^b_a R_d{}^c R_e{}^d R_f{}^e R_{hi}{}^{fg} R_{cg}{}^{hi} + 1920\,\delta^b_a R_d{}^c R_f{}^d R_g{}^e R_{hi}{}^{fg} R_{ce}{}^{hi} - 3840\,\delta^b_a R_d{}^c R_f{}^d R_h{}^e R_{ci}{}^{fg} R_{eg}{}^{hi} +\\
&+ 3840\,\delta^b_a R_d{}^c R_f{}^d R_h{}^e R_{ei}{}^{fg} R_{cg}{}^{hi} + 1920\,\delta^b_a R_f{}^c R_h{}^d R_g{}^e R_{di}{}^{fg} R_{ce}{}^{hi} + 3840\,\delta^b_a R_f{}^c R_h{}^d R_i{}^e R_{cd}{}^{fg} R_{eg}{}^{hi} - 80\,\delta^b_a R_d{}^c R_c{}^d R_{gh}{}^{ef} R_{ij}{}^{gh} R_{ef}{}^{ij} +\\
&+ 320\,\delta^b_a R_d{}^c R_c{}^d R_{gi}{}^{ef} R_{ej}{}^{gh} R_{fh}{}^{ij} - 1920\,\delta^b_a R_d{}^c R_e{}^d R_{cg}{}^{ef} R_{ij}{}^{gh} R_{fh}{}^{ij} + 960\,\delta^b_a R_d{}^c R_e{}^d R_{gh}{}^{ef} R_{ij}{}^{gh} R_{cf}{}^{ij} - 3840\,\delta^b_a R_d{}^c R_e{}^d R_{gi}{}^{ef} R_{cj}{}^{gh} R_{fh}{}^{ij} +\\
&+ 240\,\delta^b_a R_e{}^c R_f{}^d R_{cd}{}^{ef} R_{ij}{}^{gh} R_{gh}{}^{ij} - 1920\,\delta^b_a R_e{}^c R_f{}^d R_{cg}{}^{ef} R_{ij}{}^{gh} R_{dh}{}^{ij} + 480\,\delta^b_a R_e{}^c R_f{}^d R_{gh}{}^{ef} R_{ij}{}^{gh} R_{cd}{}^{ij} - 1920\,\delta^b_a R_e{}^c R_f{}^d R_{gi}{}^{ef} R_{cj}{}^{gh} R_{dh}{}^{ij} -\\
&- 1920\,\delta^b_a R_e{}^c R_g{}^d R_{cd}{}^{ef} R_{ij}{}^{gh} R_{fh}{}^{ij} - 1920\,\delta^b_a R_e{}^c R_g{}^d R_{ch}{}^{ef} R_{ij}{}^{gh} R_{df}{}^{ij} - 3840\,\delta^b_a R_e{}^c R_i{}^d R_{cg}{}^{ef} R_{dj}{}^{gh} R_{fh}{}^{ij} + 1920\,\delta^b_a R_e{}^c R_i{}^d R_{cg}{}^{ef} R_{fj}{}^{gh} R_{dh}{}^{ij} +\\
&+ 3840\,\delta^b_a R_e{}^c R_i{}^d R_{dg}{}^{ef} R_{cj}{}^{gh} R_{fh}{}^{ij} - 1920\,\delta^b_a R_e{}^c R_i{}^d R_{dg}{}^{ef} R_{fj}{}^{gh} R_{ch}{}^{ij} + 1920\,\delta^b_a R_e{}^c R_i{}^d R_{gh}{}^{ef} R_{dj}{}^{gh} R_{cf}{}^{ij} + 1920\,\delta^b_a R_e{}^c R_i{}^d R_{gj}{}^{ef} R_{cd}{}^{gh} R_{fh}{}^{ij} +\\
&+ 1920\,\delta^b_a R_d{}^c R_{cf}{}^{de} R_{eh}{}^{fg} R_{jk}{}^{hi} R_{gi}{}^{jk} - 960\,\delta^b_a R_d{}^c R_{cf}{}^{de} R_{hi}{}^{fg} R_{jk}{}^{hi} R_{eg}{}^{jk} + 3840\,\delta^b_a R_d{}^c R_{cf}{}^{de} R_{hj}{}^{fg} R_{ek}{}^{hi} R_{gi}{}^{jk} +\\
&+ 240\,\delta^b_a R_d{}^c R_{fg}{}^{de} R_{ce}{}^{fg} R_{jk}{}^{hi} R_{hi}{}^{jk} - 960\,\delta^b_a R_d{}^c R_{fg}{}^{de} R_{ch}{}^{fg} R_{jk}{}^{hi} R_{ei}{}^{jk} + 960\,\delta^b_a R_d{}^c R_{fg}{}^{de} R_{eh}{}^{fg} R_{jk}{}^{hi} R_{ci}{}^{jk} +\\
&+ 480\,\delta^b_a R_d{}^c R_{fg}{}^{de} R_{hi}{}^{fg} R_{jk}{}^{hi} R_{ce}{}^{jk} - 1920\,\delta^b_a R_d{}^c R_{fg}{}^{de} R_{hj}{}^{fg} R_{ck}{}^{hi} R_{ei}{}^{jk} - 1920\,\delta^b_a R_d{}^c R_{fh}{}^{de} R_{ce}{}^{fg} R_{jk}{}^{hi} R_{gi}{}^{jk} -\\
&- 1920\,\delta^b_a R_d{}^c R_{fh}{}^{de} R_{ci}{}^{fg} R_{jk}{}^{hi} R_{eg}{}^{jk} + 3840\,\delta^b_a R_d{}^c R_{fh}{}^{de} R_{cj}{}^{fg} R_{ek}{}^{hi} R_{gi}{}^{jk} - 3840\,\delta^b_a R_d{}^c R_{fh}{}^{de} R_{cj}{}^{fg} R_{gk}{}^{hi} R_{ei}{}^{jk} +\\
&+ 1920\,\delta^b_a R_d{}^c R_{fh}{}^{de} R_{ei}{}^{fg} R_{jk}{}^{hi} R_{cg}{}^{jk} + 3840\,\delta^b_a R_d{}^c R_{fh}{}^{de} R_{ej}{}^{fg} R_{gk}{}^{hi} R_{ci}{}^{jk} - 1920\,\delta^b_a R_d{}^c R_{fh}{}^{de} R_{ij}{}^{fg} R_{gk}{}^{hi} R_{ce}{}^{jk} +\\
&+ 20\,\delta^b_a R_{ef}{}^{cd} R_{cd}{}^{ef} R_{ij}{}^{gh} R_{kl}{}^{ij} R_{gh}{}^{kl} - 80\,\delta^b_a R_{ef}{}^{cd} R_{cd}{}^{ef} R_{ik}{}^{gh} R_{gl}{}^{ij} R_{hj}{}^{kl} + 480\,\delta^b_a R_{ef}{}^{cd} R_{cg}{}^{ef} R_{di}{}^{gh} R_{kl}{}^{ij} R_{hj}{}^{kl} -\\
&- 480\,\delta^b_a R_{ef}{}^{cd} R_{cg}{}^{ef} R_{ij}{}^{gh} R_{kl}{}^{ij} R_{dh}{}^{kl} + 1920\,\delta^b_a R_{ef}{}^{cd} R_{cg}{}^{ef} R_{ik}{}^{gh} R_{dl}{}^{ij} R_{hj}{}^{kl} + 24\,\delta^b_a R_{ef}{}^{cd} R_{gh}{}^{ef} R_{ij}{}^{gh} R_{kl}{}^{ij} R_{cd}{}^{kl} -\\
&- 480\,\delta^b_a R_{ef}{}^{cd} R_{gh}{}^{ef} R_{ik}{}^{gh} R_{cl}{}^{ij} R_{dj}{}^{kl} - 480\,\delta^b_a R_{ef}{}^{cd} R_{gi}{}^{ef} R_{cj}{}^{gh} R_{kl}{}^{ij} R_{dh}{}^{kl} + 960\,\delta^b_a R_{ef}{}^{cd} R_{gi}{}^{ef} R_{ck}{}^{gh} R_{dl}{}^{ij} R_{hj}{}^{kl} -\\
&- 1920\,\delta^b_a R_{ef}{}^{cd} R_{gi}{}^{ef} R_{ck}{}^{gh} R_{hl}{}^{ij} R_{dj}{}^{kl} + 1920\,\delta^b_a R_{eg}{}^{cd} R_{ch}{}^{ef} R_{ik}{}^{gh} R_{dl}{}^{ij} R_{fj}{}^{kl} - 384\,\delta^b_a R_{eg}{}^{cd} R_{ci}{}^{ef} R_{dk}{}^{gh} R_{fl}{}^{ij} R_{hj}{}^{kl} +\\
&+ 1920\,\delta^b_a R_{eg}{}^{cd} R_{ci}{}^{ef} R_{dk}{}^{gh} R_{hl}{}^{ij} R_{fj}{}^{kl} - 1920\,\delta^b_a R_{eg}{}^{cd} R_{ci}{}^{ef} R_{fk}{}^{gh} R_{hl}{}^{ij} R_{dj}{}^{kl} - 768\,\delta^b_a R_{eg}{}^{cd} R_{hi}{}^{ef} R_{jk}{}^{gh} R_{cl}{}^{ij} R_{df}{}^{kl} + 10\,R_a{}^b R^4 - 80\,R_a{}^c R^3 R_c{}^b -\\
&- 240\,R_a{}^b R^2 R_d{}^c R_c{}^d + 480\,R_a{}^c R^2 R_d{}^b R_c{}^d - 480\,R_a{}^c R^2 R_e{}^d R_{cd}{}^{be} + 60\,R_a{}^b R^2 R_{ef}{}^{cd} R_{cd}{}^{ef} - 240\,R_a{}^c R^2 R_{ef}{}^{bd} R_{cd}{}^{ef} + 640\,R_a{}^b R R_d{}^c R_e{}^d R_c{}^e +\\
&+ 960\,R_a{}^c R R_c{}^b R_e{}^d R_d{}^e - 1920\,R_a{}^c R R_d{}^b R_e{}^d R_c{}^e - 960\,R_a{}^b R R_e{}^c R_f{}^d R_{cd}{}^{ef} + 1920\,R_a{}^c R R_e{}^b R_f{}^d R_{cd}{}^{ef} + 1920\,R_a{}^c R R_e{}^d R_f{}^e R_{cd}{}^{bf} -\\
&- 1920\,R_a{}^c R R_f{}^d R_c{}^e R_{de}{}^{bf} - 960\,R_a{}^b R R_d{}^c R_{fg}{}^{de} R_{ce}{}^{fg} - 240\,R_a{}^c R R_c{}^b R_{fg}{}^{de} R_{de}{}^{fg} + 960\,R_a{}^c R R_d{}^b R_{fg}{}^{de} R_{ce}{}^{fg} + 960\,R_a{}^c R R_c{}^d R_{fg}{}^{be} R_{de}{}^{fg} +\\
&+ 960\,R_a{}^c R R_e{}^d R_{fg}{}^{be} R_{cd}{}^{fg} - 1920\,R_a{}^c R R_f{}^d R_{cg}{}^{be} R_{de}{}^{fg} + 1920\,R_a{}^c R R_f{}^d R_{dg}{}^{be} R_{ce}{}^{fg} - 80\,R_a{}^b R R_{ef}{}^{cd} R_{gh}{}^{ef} R_{cd}{}^{gh} +\\
&+ 320\,R_a{}^b R R_{eg}{}^{cd} R_{ch}{}^{ef} R_{df}{}^{gh} - 960\,R_a{}^c R R_{ce}{}^{bd} R_{gh}{}^{ef} R_{df}{}^{gh} + 480\,R_a{}^c R R_{ef}{}^{bd} R_{gh}{}^{ef} R_{cd}{}^{gh} + 1920\,R_a{}^c R R_{eg}{}^{bd} R_{dh}{}^{ef} R_{cf}{}^{gh} +\\
&+ 480\,R_a{}^b R_d{}^c R_c{}^d R_f{}^e R_e{}^f - 960\,R_a{}^b R_d{}^c R_e{}^d R_f{}^e R_c{}^f - 1280\,R_a{}^c R_c{}^b R_e{}^d R_f{}^e R_d{}^f - 1920\,R_a{}^c R_d{}^b R_c{}^d R_f{}^e R_e{}^f + 3840\,R_a{}^c R_d{}^b R_e{}^d R_f{}^e R_c{}^f +\\
&+ 3840\,R_a{}^b R_d{}^c R_f{}^d R_g{}^e R_{ce}{}^{fg} + 1920\,R_a{}^c R_c{}^b R_f{}^d R_g{}^e R_{de}{}^{fg} - 3840\,R_a{}^c R_d{}^b R_f{}^d R_g{}^e R_{ce}{}^{fg} - 3840\,R_a{}^c R_f{}^b R_e{}^d R_g{}^e R_{cd}{}^{fg} +
\end{aligned}$$

$$+ 3840 R_a^c R_f^b R_g^d R_c^e R_{de}^{fg} - 3840 R_a^c R_c^d R_f^e R_g^f R_{de}^{bg} - 3840 R_a^c R_e^d R_f^e R_g^f R_{cd}^{bg} + 3840 R_a^c R_g^d R_f^e R_c^f R_{de}^{bg} +$$

$$+ 1920 R_a^c R_g^d R_f^e R_e^f R_{cd}^{bg} - 240 R_a^b R_d^c R_c^d R_{gh}^{ef} R_{ef}^{gh} + 1920 R_a^b R_d^c R_e^d R_{gh}^{ef} R_{cf}^{gh} + 960 R_a^b R_e^c R_f^d R_{gh}^{ef} R_{cd}^{gh} -$$

$$- 1920 R_a^b R_e^c R_g^d R_{ch}^{ef} R_{df}^{gh} + 1920 R_a^b R_e^c R_g^d R_{dh}^{ef} R_{cf}^{gh} + 1920 R_a^c R_c^b R_e^d R_{gh}^{ef} R_{df}^{gh} + 480 R_a^c R_d^b R_c^d R_{gh}^{ef} R_{ef}^{gh} -$$

$$- 1920 R_a^c R_d^b R_e^d R_{gh}^{ef} R_{cf}^{gh} - 1920 R_a^c R_e^b R_c^d R_{gh}^{ef} R_{df}^{gh} - 1920 R_a^c R_e^b R_f^d R_{gh}^{ef} R_{cd}^{gh} + 3840 R_a^c R_e^b R_g^d R_{ch}^{ef} R_{df}^{gh} -$$

$$- 3840 R_a^c R_e^b R_g^d R_{dh}^{ef} R_{cf}^{gh} + 3840 R_a^c R_c^d R_g^e R_{dh}^{bf} R_{ef}^{gh} - 1920 R_a^c R_e^d R_c^e R_{gh}^{bf} R_{df}^{gh} + 960 R_a^c R_e^d R_d^e R_{gh}^{bf} R_{cf}^{gh} -$$

$$- 1920 R_a^c R_e^d R_f^e R_{gh}^{bf} R_{cd}^{gh} + 3840 R_a^c R_e^d R_g^e R_{ch}^{bf} R_{df}^{gh} - 3840 R_a^c R_e^d R_g^e R_{dh}^{bf} R_{cf}^{gh} + 1920 R_a^c R_f^d R_c^e R_{gh}^{bf} R_{de}^{gh} -$$

$$- 3840 R_a^c R_f^d R_g^e R_{ch}^{bf} R_{de}^{gh} + 3840 R_a^c R_f^d R_g^e R_{dh}^{bf} R_{ce}^{gh} - 3840 R_a^c R_g^d R_c^e R_{dh}^{bf} R_{ef}^{gh} - 3840 R_a^c R_g^d R_f^e R_{dh}^{bf} R_{ce}^{gh} -$$

$$- 3840 R_a^c R_g^d R_h^e R_{cd}^{bf} R_{ef}^{gh} - 1920 R_a^c R_g^d R_h^e R_{de}^{bf} R_{cf}^{gh} - 1920 R_a^b R_d^c R_{cf}^{de} R_{hi}^{fg} R_{eg}^{hi} + 960 R_a^b R_d^c R_{fg}^{de} R_{hi}^{fg} R_{ce}^{hi} -$$

$$- 3840 R_a^b R_d^c R_{fh}^{de} R_{ci}^{fg} R_{eg}^{hi} + 160 R_a^c R_c^b R_{fg}^{de} R_{hi}^{fg} R_{de}^{hi} - 640 R_a^c R_c^b R_{fh}^{de} R_{di}^{fg} R_{eg}^{hi} + 1920 R_a^c R_d^b R_{cf}^{de} R_{hi}^{fg} R_{eg}^{hi} -$$

$$- 960 R_a^c R_d^b R_{fg}^{de} R_{hi}^{fg} R_{ce}^{hi} - 3840 R_a^c R_d^b R_{fh}^{de} R_{ei}^{fg} R_{cg}^{hi} + 1920 R_a^c R_c^d R_{df}^{be} R_{hi}^{fg} R_{eg}^{hi} - 960 R_a^c R_c^d R_{fg}^{be} R_{hi}^{fg} R_{de}^{hi} -$$

$$- 3840 R_a^c R_c^d R_{fh}^{be} R_{ei}^{fg} R_{dg}^{hi} - 480 R_a^c R_e^d R_{cd}^{be} R_{hi}^{fg} R_{fg}^{hi} + 1920 R_a^c R_e^d R_{cf}^{be} R_{hi}^{fg} R_{dg}^{hi} - 1920 R_a^c R_e^d R_{df}^{be} R_{hi}^{fg} R_{cg}^{hi} -$$

$$- 960 R_a^c R_e^d R_{fg}^{be} R_{hi}^{fg} R_{cd}^{hi} + 3840 R_a^c R_e^d R_{fh}^{be} R_{ci}^{fg} R_{dg}^{hi} + 1920 R_a^c R_f^d R_{cd}^{be} R_{hi}^{fg} R_{eg}^{hi} + 1920 R_a^c R_f^d R_{cg}^{be} R_{hi}^{fg} R_{de}^{hi} -$$

$$- 1920 R_a^c R_f^d R_{dg}^{be} R_{hi}^{fg} R_{ce}^{hi} + 3840 R_a^c R_f^d R_{gh}^{be} R_{ei}^{fg} R_{cd}^{hi} + 3840 R_a^c R_h^d R_{cf}^{be} R_{di}^{fg} R_{eg}^{hi} - 3840 R_a^c R_h^d R_{cf}^{be} R_{ei}^{fg} R_{dg}^{hi} -$$

$$- 3840 R_a^c R_h^d R_{df}^{be} R_{ci}^{fg} R_{eg}^{hi} + 3840 R_a^c R_h^d R_{df}^{be} R_{ei}^{fg} R_{cg}^{hi} + 1920 R_a^c R_h^d R_{fg}^{be} R_{ci}^{fg} R_{de}^{hi} - 1920 R_a^c R_h^d R_{fg}^{be} R_{di}^{fg} R_{ce}^{hi} +$$

$$+ 1920 R_a^c R_h^d R_{fg}^{be} R_{ei}^{fg} R_{cd}^{hi} + 3840 R_a^c R_h^d R_{fi}^{be} R_{ce}^{fg} R_{dg}^{hi} - 3840 R_a^c R_h^d R_{fi}^{be} R_{de}^{fg} R_{cg}^{hi} + 30 R_a^b R_{ef}^{cd} R_{cd}^{ef} R_{ij}^{gh} R_{gh}^{ij} -$$

$$- 480 R_a^b R_{ef}^{cd} R_{cg}^{ef} R_{ij}^{gh} R_{dh}^{ij} + 60 R_a^b R_{ef}^{cd} R_{gh}^{ef} R_{ij}^{gh} R_{cd}^{ij} - 960 R_a^b R_{ef}^{cd} R_{gi}^{ef} R_{cj}^{gh} R_{dh}^{ij} + 480 R_a^b R_{eg}^{cd} R_{ci}^{ef} R_{dj}^{gh} R_{fh}^{ij} -$$

$$- 960 R_a^b R_{eg}^{cd} R_{ci}^{ef} R_{fj}^{gh} R_{dh}^{ij} - 1920 R_a^c R_{ce}^{bd} R_{dg}^{ef} R_{ij}^{gh} R_{fh}^{ij} + 960 R_a^c R_{ce}^{bd} R_{gh}^{ef} R_{ij}^{gh} R_{df}^{ij} - 3840 R_a^c R_{ce}^{bd} R_{gi}^{ef} R_{dj}^{gh} R_{fh}^{ij} -$$

$$- 240 R_a^c R_{ef}^{bd} R_{cd}^{ef} R_{ij}^{gh} R_{gh}^{ij} + 960 R_a^c R_{ef}^{bd} R_{cg}^{ef} R_{ij}^{gh} R_{dh}^{ij} - 960 R_a^c R_{ef}^{bd} R_{dg}^{ef} R_{ij}^{gh} R_{ch}^{ij} - 480 R_a^c R_{ef}^{bd} R_{gh}^{ef} R_{ij}^{gh} R_{cd}^{ij} -$$

$$- 1920 R_a^c R_{ef}^{bd} R_{gi}^{ef} R_{dj}^{gh} R_{ch}^{ij} + 1920 R_a^c R_{eg}^{bd} R_{cd}^{ef} R_{ij}^{gh} R_{fh}^{ij} + 1920 R_a^c R_{eg}^{bd} R_{ch}^{ef} R_{ij}^{gh} R_{df}^{ij} - 1920 R_a^c R_{eg}^{bd} R_{dh}^{ef} R_{ij}^{gh} R_{cf}^{ij} +$$

$$+ 3840 R_a^c R_{eg}^{bd} R_{di}^{ef} R_{cj}^{gh} R_{fh}^{ij} - 3840 R_a^c R_{eg}^{bd} R_{di}^{ef} R_{fj}^{gh} R_{ch}^{ij} + 3840 R_a^c R_{eg}^{bd} R_{hi}^{ef} R_{cj}^{gh} R_{df}^{ij} + 1920 R_a^c R_{eg}^{bd} R_{hi}^{ef} R_{fj}^{gh} R_{cd}^{ij} +$$

$$+ 80 R_{ad}^{bc} R^3 R_c^d + 40 R_{ae}^{cd} R^3 R_{cd}^{be} - 480 R_{ad}^{bc} R^2 R_e^d R_c^e - 480 R_{ae}^{cd} R^2 R_c^b R_d^e + 480 R_{ae}^{bc} R^2 R_f^d R_{cd}^{ef} + 240 R_{ae}^{cd} R^2 R_f^b R_{cd}^{ef} -$$

$$- 240 R_{ae}^{cd} R^2 R_f^e R_{cd}^{bf} + 480 R_{af}^{cd} R^2 R_c^e R_{de}^{bf} + 240 R_{ad}^{bc} R^2 R_{fg}^{de} R_{ce}^{fg} - 120 R_{ae}^{cd} R^2 R_{fg}^{be} R_{cd}^{fg} + 480 R_{af}^{cd} R^2 R_{cg}^{be} R_{de}^{fg} -$$

$$- 960 R_{ad}^{bc} R R_c^d R_f^e R_e^f + 1920 R_{ad}^{bc} R R_e^d R_f^e R_c^f + 1920 R_{ae}^{cd} R R_c^b R_f^e R_d^f + 1920 R_{af}^{cd} R R_e^b R_c^e R_d^f - 1920 R_{ad}^{bc} R R_f^d R_g^e R_{ce}^{fg} -$$

$$- 1920 R_{af}^{bc} R R_c^d R_g^e R_{de}^{fg} - 1920 R_{af}^{bc} R R_e^d R_g^e R_{cd}^{fg} + 960 R_{ae}^{cd} R R_f^b R_g^e R_{cd}^{fg} - 1920 R_{af}^{cd} R R_c^b R_g^e R_{de}^{fg} -$$

$$- 960 R_{af}^{cd} R R_e^b R_g^e R_{cd}^{fg} + 1920 R_{af}^{cd} R R_g^b R_c^e R_{de}^{fg} + 960 R_{ae}^{cd} R R_f^e R_g^f R_{cd}^{bg} - 1920 R_{af}^{cd} R R_c^e R_g^f R_{de}^{bg} +$$

$$+ 1920 R_{af}^{cd} R R_g^e R_c^f R_{de}^{bg} + 960 R_{ag}^{cd} R R_c^e R_d^f R_{ef}^{bg} - 1920 R_{ag}^{cd} R R_f^e R_c^f R_{de}^{bg} - 480 R_{ag}^{cd} R R_f^e R_e^f R_{cd}^{bg} +$$

$$+ 240 R_{ad}^{bc} R R_c^d R_{gh}^{ef} R_{ef}^{gh} - 960 R_{ad}^{bc} R R_e^d R_{gh}^{ef} R_{cf}^{gh} - 960 R_{ae}^{bc} R R_c^d R_{gh}^{ef} R_{df}^{gh} - 960 R_{ae}^{bc} R R_f^d R_{gh}^{ef} R_{cd}^{gh} +$$

$$+ 1920 R_{ae}^{bc} R R_g^d R_{ch}^{ef} R_{df}^{gh} - 1920 R_{ae}^{bc} R R_g^d R_{dh}^{ef} R_{cf}^{gh} - 960 R_{ae}^{cd} R R_c^b R_{gh}^{ef} R_{df}^{gh} - 480 R_{ae}^{cd} R R_f^b R_{gh}^{ef} R_{cd}^{gh} -$$

$$- 1920 R_{ag}^{cd} R R_e^b R_{ch}^{ef} R_{df}^{gh} + 960 R_{ae}^{cd} R R_c^e R_{gh}^{bf} R_{df}^{gh} + 480 R_{ae}^{cd} R R_f^e R_{gh}^{bf} R_{cd}^{gh} - 1920 R_{ae}^{cd} R R_g^e R_{ch}^{bf} R_{df}^{gh} -$$

$$- 960 R_{af}^{cd} R R_c^e R_{gh}^{bf} R_{de}^{gh} + 1920 R_{af}^{cd} R R_g^e R_{ch}^{bf} R_{de}^{gh} + 960 R_{af}^{cd} R R_g^e R_{eh}^{bf} R_{cd}^{gh} + 1920 R_{ag}^{cd} R R_c^e R_{dh}^{bf} R_{ef}^{gh} -$$

$$- 1920 R_{ag}^{cd} R R_c^e R_{eh}^{bf} R_{df}^{gh} - 1920 R_{ag}^{cd} R R_f^e R_{ch}^{bf} R_{de}^{gh} - 960 R_{ag}^{cd} R R_f^e R_{eh}^{bf} R_{cd}^{gh} + 960 R_{ag}^{cd} R R_h^e R_{cd}^{bf} R_{ef}^{gh} -$$

$$- 1920 R_{ag}^{cd} R R_h^e R_{ce}^{bf} R_{df}^{gh} + 960 R_{ad}^{bc} R R_{cf}^{de} R_{hi}^{fg} R_{eg}^{hi} - 480 R_{ad}^{bc} R R_{fg}^{de} R_{hi}^{fg} R_{ce}^{hi} + 1920 R_{ad}^{bc} R R_{fh}^{de} R_{ci}^{fg} R_{eg}^{hi} +$$

$$+ 120 R_{ae}^{cd} R R_{cd}^{be} R_{hi}^{fg} R_{fg}^{hi} - 960 R_{ae}^{cd} R R_{cf}^{be} R_{hi}^{fg} R_{dg}^{hi} + 240 R_{ae}^{cd} R R_{fg}^{be} R_{hi}^{fg} R_{cd}^{hi} - 960 R_{ae}^{cd} R R_{fh}^{be} R_{ci}^{fg} R_{dg}^{hi} -$$

$$- 480 R_{af}^{cd} R R_{cd}^{be} R_{hi}^{fg} R_{eg}^{hi} - 960 R_{af}^{cd} R R_{cg}^{be} R_{hi}^{fg} R_{de}^{hi} - 960 R_{af}^{cd} R R_{gh}^{be} R_{ei}^{fg} R_{cd}^{hi} - 1920 R_{ah}^{cd} R R_{cf}^{be} R_{di}^{fg} R_{eg}^{hi} +$$

$$+ 1920 R_{ah}^{cd} R R_{cf}^{be} R_{ei}^{fg} R_{dg}^{hi} - 960 R_{ah}^{cd} R R_{fg}^{be} R_{ci}^{fg} R_{de}^{hi} - 480 R_{ah}^{cd} R R_{fg}^{be} R_{ei}^{fg} R_{cd}^{hi} - 1920 R_{ah}^{cd} R R_{fi}^{be} R_{ce}^{fg} R_{dg}^{hi} +$$

$$+ 1280 R_{ad}^{bc} R_c^d R_f^e R_g^f R_e^g + 1920 R_{ad}^{bc} R_e^d R_c^e R_g^f R_f^g - 3840 R_{ad}^{bc} R_e^d R_f^e R_g^f R_c^g + 1920 R_{ae}^{cd} R_c^b R_d^e R_g^f R_f^g -$$

$$- 3840 R_{ae}^{cd} R_c^b R_f^e R_g^f R_d^g - 3840 R_{af}^{cd} R_e^b R_c^e R_g^f R_d^g - 3840 R_{ag}^{cd} R_e^b R_f^e R_c^f R_d^g - 1920 R_{ad}^{bc} R_c^d R_g^e R_h^f R_{ef}^{gh} +$$

$$+ 3840 R_{ad}^{bc} R_e^d R_g^e R_h^f R_{cf}^{gh} + 3840 R_{ad}^{bc} R_g^d R_c^e R_h^f R_{ef}^{gh} + 3840 R_{ad}^{bc} R_g^d R_f^e R_h^f R_{ce}^{gh} + 3840 R_{ag}^{bc} R_c^d R_f^e R_h^f R_{de}^{gh} +$$

$$+ 3840 R_{ag}^{bc} R_e^d R_f^e R_h^f R_{cd}^{gh} - 3840 R_{ag}^{bc} R_h^d R_f^e R_c^f R_{de}^{gh} - 1920 R_{ag}^{bc} R_h^d R_f^e R_e^f R_{cd}^{gh} + 3840 R_{ae}^{cd} R_c^b R_g^e R_h^f R_{df}^{gh} -$$

$$- 1920 R_{ae}^{cd} R_g^b R_f^e R_h^f R_{cd}^{gh} - 1920 R_{af}^{cd} R_e^b R_g^e R_h^f R_{cd}^{gh} + 3840 R_{af}^{cd} R_g^b R_c^e R_h^f R_{de}^{gh} - 3840 R_{af}^{cd} R_g^b R_h^e R_c^f R_{de}^{gh} +$$

$$+ 3840 R_{ag}^{cd} R_c^b R_d^e R_h^f R_{ef}^{gh} + 3840 R_{ag}^{cd} R_c^b R_f^e R_h^f R_{de}^{gh} + 3840 R_{ag}^{cd} R_e^b R_c^e R_h^f R_{df}^{gh} + 1920 R_{ag}^{cd} R_e^b R_f^e R_h^f R_{cd}^{gh} -$$

$$- 3840 R_{ag}^{cd} R_e^b R_h^e R_c^f R_{df}^{gh} + 1920 R_{ag}^{cd} R_h^b R_c^e R_d^f R_{ef}^{gh} - 3840 R_{ag}^{cd} R_h^b R_f^e R_c^f R_{de}^{gh} - 960 R_{ag}^{cd} R_h^b R_f^e R_e^f R_{cd}^{gh} -$$

$$- 1920 R_{ae}^{cd} R_f^e R_g^f R_h^g R_{cd}^{bh} + 960 R_{ae}^{cd} R_h^e R_g^f R_f^g R_{cd}^{bh} + 3840 R_{af}^{cd} R_c^e R_g^f R_h^g R_{de}^{bh} - 3840 R_{af}^{cd} R_g^e R_c^f R_h^g R_{de}^{bh} +$$

$$+ 3840 R_{af}^{cd} R_g^e R_h^f R_c^g R_{de}^{bh} - 3840 R_{af}^{cd} R_h^e R_g^f R_c^g R_{de}^{bh} - 1920 R_{ag}^{cd} R_c^e R_d^f R_h^g R_{ef}^{bh} - 3840 R_{ag}^{cd} R_h^e R_c^f R_d^g R_{ef}^{bh} -$$

$$- 3840 R_{ah}^{cd} R_c^e R_g^f R_d^g R_{ef}^{bh} - 1920 R_{ah}^{cd} R_c^e R_g^f R_f^g R_{de}^{bh} + 3840 R_{ah}^{cd} R_f^e R_g^f R_c^g R_{de}^{bh} + 640 R_{ah}^{cd} R_f^e R_g^f R_e^g R_{cd}^{bh} -$$

$$- 1920 R_{ad}^{bc} R_c^d R_f^e R_{hi}^{fg} R_{eg}^{hi} - 480 R_{ad}^{bc} R_e^d R_c^e R_{hi}^{fg} R_{fg}^{hi} + 1920 R_{ad}^{bc} R_e^d R_f^e R_{hi}^{fg} R_{cg}^{hi} + 1920 R_{ad}^{bc} R_f^d R_c^e R_{hi}^{fg} R_{eg}^{hi} +$$

$$+ 1920 R_{ad}^{bc} R_f^d R_g^e R_{hi}^{fg} R_{ce}^{hi} - 3840 R_{ad}^{bc} R_f^d R_h^e R_{ci}^{fg} R_{eg}^{hi} + 3840 R_{ad}^{bc} R_f^d R_h^e R_{ei}^{fg} R_{cg}^{hi} + 1920 R_{af}^{bc} R_c^d R_g^e R_{hi}^{fg} R_{de}^{hi} -$$

$$- 3840 R_{af}^{bc} R_c^d R_h^e R_{di}^{fg} R_{eg}^{hi} + 1920 R_{af}^{bc} R_e^d R_c^e R_{hi}^{fg} R_{dg}^{hi} - 960 R_{af}^{bc} R_e^d R_d^e R_{hi}^{fg} R_{cg}^{hi} + 1920 R_{af}^{bc} R_e^d R_g^e R_{hi}^{fg} R_{cd}^{hi} -$$

$$- 3840 R_{af}^{bc} R_e^d R_h^e R_{ci}^{fg} R_{dg}^{hi} + 3840 R_{af}^{bc} R_e^d R_h^e R_{di}^{fg} R_{cg}^{hi} + 3840 R_{af}^{bc} R_g^d R_h^e R_{ci}^{fg} R_{de}^{hi} - 3840 R_{af}^{bc} R_g^d R_h^e R_{di}^{fg} R_{ce}^{hi} +$$

$$+ 3840 R_{af}^{bc} R_h^d R_c^e R_{di}^{fg} R_{eg}^{hi} + 3840 R_{af}^{bc} R_h^d R_g^e R_{di}^{fg} R_{ce}^{hi} + 3840 R_{af}^{bc} R_h^d R_i^e R_{cd}^{fg} R_{eg}^{hi} + 1920 R_{af}^{bc} R_h^d R_i^e R_{de}^{fg} R_{cg}^{hi} -$$

$$- 480 R_{ae}^{cd} R_c^b R_d^e R_{hi}^{fg} R_{fg}^{hi} + 1920 R_{ae}^{cd} R_c^b R_f^e R_{hi}^{fg} R_{dg}^{hi} - 1920 R_{ae}^{cd} R_f^b R_c^e R_{hi}^{fg} R_{dg}^{hi} - 960 R_{ae}^{cd} R_f^b R_g^e R_{hi}^{fg} R_{cd}^{hi} +$$

$$+ 3840 R_{ae}^{cd} R_f^b R_h^e R_{ci}^{fg} R_{dg}^{hi} + 1920 R_{af}^{cd} R_c^b R_d^e R_{hi}^{fg} R_{eg}^{hi} + 1920 R_{af}^{cd} R_c^b R_g^e R_{hi}^{fg} R_{de}^{hi} - 3840 R_{af}^{cd} R_c^b R_h^e R_{di}^{fg} R_{eg}^{hi} +$$

$$+ 3840 R_{af}^{cd} R_c^b R_h^e R_{ei}^{fg} R_{dg}^{hi} + 1920 R_{af}^{cd} R_e^b R_c^e R_{hi}^{fg} R_{dg}^{hi} + 960 R_{af}^{cd} R_e^b R_g^e R_{hi}^{fg} R_{cd}^{hi} - 1920 R_{af}^{cd} R_g^b R_c^e R_{hi}^{fg} R_{de}^{hi} +$$

$$+ 3840 R_{af}^{cd} R_g^b R_h^e R_{ci}^{fg} R_{de}^{hi} + 1920 R_{af}^{cd} R_g^b R_h^e R_{ei}^{fg} R_{cd}^{hi} + 3840 R_{ah}^{cd} R_e^b R_f^e R_{ci}^{fg} R_{dg}^{hi} - 3840 R_{ah}^{cd} R_f^b R_c^e R_{di}^{fg} R_{eg}^{hi} +$$

$$+ 3840 R_{ah}^{cd} R_f^b R_c^e R_{ei}^{fg} R_{dg}^{hi} + 3840 R_{ah}^{cd} R_f^b R_g^e R_{ci}^{fg} R_{de}^{hi} + 1920 R_{ah}^{cd} R_f^b R_g^e R_{ei}^{fg} R_{cd}^{hi} - 1920 R_{ah}^{cd} R_f^b R_i^e R_{cd}^{fg} R_{eg}^{hi} +$$

$$+ 3840 R_{ah}^{cd} R_f^b R_i^e R_{ce}^{fg} R_{dg}^{hi} + 3840 R_{ae}^{cd} R_c^e R_h^f R_{di}^{bg} R_{fg}^{hi} - 1920 R_{ae}^{cd} R_f^e R_c^f R_{hi}^{bg} R_{dg}^{hi} - 960 R_{ae}^{cd} R_f^e R_g^f R_{hi}^{bg} R_{cd}^{hi} +$$

$$+ 3840 R_{ae}^{cd} R_f^e R_h^f R_{ci}^{bg} R_{dg}^{hi} - 3840 R_{ae}^{cd} R_g^e R_h^f R_{ci}^{bg} R_{df}^{hi} - 3840 R_{ae}^{cd} R_h^e R_c^f R_{di}^{bg} R_{fg}^{hi} + 3840 R_{ae}^{cd} R_h^e R_g^f R_{ci}^{bg} R_{df}^{hi} -$$

$$- 1920 R_{ae}^{cd} R_h^e R_i^f R_{cd}^{bg} R_{fg}^{hi} + 1920 R_{af}^{cd} R_c^e R_d^f R_{hi}^{bg} R_{eg}^{hi} + 1920 R_{af}^{cd} R_c^e R_g^f R_{hi}^{bg} R_{de}^{hi} + 3840 R_{af}^{cd} R_c^e R_h^f R_{ei}^{bg} R_{dg}^{hi} -$$

$$- 1920 R_{af}^{cd} R_g^e R_c^f R_{hi}^{bg} R_{de}^{hi} + 1920 R_{af}^{cd} R_g^e R_h^f R_{ei}^{bg} R_{cd}^{hi} - 3840 R_{af}^{cd} R_h^e R_c^f R_{ei}^{bg} R_{dg}^{hi} - 1920 R_{af}^{cd} R_h^e R_g^f R_{ei}^{bg} R_{cd}^{hi} -$$

$$- 3840 R_{af}^{cd} R_h^e R_i^f R_{ce}^{bg} R_{dg}^{hi} - 960 R_{ag}^{cd} R_c^e R_d^f R_{hi}^{bg} R_{ef}^{hi} + 3840 R_{ag}^{cd} R_c^e R_h^f R_{di}^{bg} R_{ef}^{hi} - 3840 R_{ag}^{cd} R_c^e R_h^f R_{ei}^{bg} R_{df}^{hi} +$$

$$+ 1920 R_{ag}^{cd} R_f^e R_c^f R_{hi}^{bg} R_{de}^{hi} + 480 R_{ag}^{cd} R_f^e R_e^f R_{hi}^{bg} R_{cd}^{hi} - 3840 R_{ag}^{cd} R_f^e R_h^f R_{ci}^{bg} R_{de}^{hi} - 1920 R_{ag}^{cd} R_f^e R_h^f R_{ei}^{bg} R_{cd}^{hi} +$$

$$+ 3840 R_{ag}^{cd} R_h^e R_c^f R_{ei}^{bg} R_{df}^{hi} - 960 R_{ag}^{cd} R_h^e R_i^f R_{cd}^{bg} R_{ef}^{hi} + 3840 R_{ag}^{cd} R_h^e R_i^f R_{ce}^{bg} R_{df}^{hi} - 960 R_{ag}^{cd} R_h^e R_i^f R_{ef}^{bg} R_{cd}^{hi} +$$

$$+ 3840 R_{ah}^{cd} R_c^e R_d^f R_{ei}^{bg} R_{fg}^{hi} + 3840 R_{ah}^{cd} R_c^e R_g^f R_{ei}^{bg} R_{df}^{hi} + 3840 R_{ah}^{cd} R_c^e R_i^f R_{de}^{bg} R_{fg}^{hi} + 3840 R_{ah}^{cd} R_c^e R_i^f R_{ef}^{bg} R_{dg}^{hi} -$$

$$- 3840 R_{ah}^{cd} R_f^e R_c^f R_{di}^{bg} R_{eg}^{hi} + 3840 R_{ah}^{cd} R_f^e R_c^f R_{ei}^{bg} R_{dg}^{hi} - 1920 R_{ah}^{cd} R_f^e R_e^f R_{ci}^{bg} R_{dg}^{hi} + 3840 R_{ah}^{cd} R_f^e R_g^f R_{ci}^{bg} R_{de}^{hi} +$$

$$+ 1920 R_{ah}^{cd} R_f^e R_g^f R_{ei}^{bg} R_{cd}^{hi} - 1920 R_{ah}^{cd} R_f^e R_i^f R_{cd}^{bg} R_{eg}^{hi} + 3840 R_{ah}^{cd} R_f^e R_i^f R_{ce}^{bg} R_{dg}^{hi} + 3840 R_{ah}^{cd} R_g^e R_c^f R_{di}^{bg} R_{ef}^{hi} -$$

$$- 3840 R_{ah}^{cd} R_g^e R_c^f R_{ei}^{bg} R_{df}^{hi} + 1920 R_{ah}^{cd} R_g^e R_i^f R_{cd}^{bg} R_{ef}^{hi} - 3840 R_{ah}^{cd} R_g^e R_i^f R_{ce}^{bg} R_{df}^{hi} + 1920 R_{ah}^{cd} R_g^e R_i^f R_{ef}^{bg} R_{cd}^{hi} -$$

$$- 3840 R_{ah}^{cd} R_i^e R_c^f R_{de}^{bg} R_{fg}^{hi} + 3840 R_{ah}^{cd} R_i^e R_g^f R_{ce}^{bg} R_{df}^{hi} - 160 R_{ad}^{bc} R_c^d R_{gh}^{ef} R_{ij}^{gh} R_{ef}^{ij} + 640 R_{ad}^{bc} R_c^d R_{gi}^{ef} R_{ej}^{gh} R_{fh}^{ij} -$$

$$- 1920 R_{ad}^{bc} R_e^d R_{cg}^{ef} R_{ij}^{gh} R_{fh}^{ij} + 960 R_{ad}^{bc} R_e^d R_{gh}^{ef} R_{ij}^{gh} R_{cf}^{ij} - 3840 R_{ad}^{bc} R_e^d R_{gi}^{ef} R_{cj}^{gh} R_{fh}^{ij} - 1920 R_{ae}^{bc} R_c^d R_{dg}^{ef} R_{ij}^{gh} R_{fh}^{ij} +$$

$$+ 960 R_{ae}^{bc} R_c^d R_{gh}^{ef} R_{ij}^{gh} R_{df}^{ij} + 3840 R_{ae}^{bc} R_c^d R_{gi}^{ef} R_{fj}^{gh} R_{dh}^{ij} + 480 R_{ae}^{bc} R_f^d R_{cd}^{ef} R_{ij}^{gh} R_{gh}^{ij} - 1920 R_{ae}^{bc} R_f^d R_{cg}^{ef} R_{ij}^{gh} R_{dh}^{ij} +$$

$$+ 1920 R_{ae}^{bc} R_f^d R_{dg}^{ef} R_{ij}^{gh} R_{ch}^{ij} + 960 R_{ae}^{bc} R_f^d R_{gh}^{ef} R_{ij}^{gh} R_{cd}^{ij} - 3840 R_{ae}^{bc} R_f^d R_{gi}^{ef} R_{cj}^{gh} R_{dh}^{ij} - 1920 R_{ae}^{bc} R_g^d R_{cd}^{ef} R_{ij}^{gh} R_{fh}^{ij} -$$

$$- 1920 R_{ae}^{bc} R_g^d R_{ch}^{ef} R_{ij}^{gh} R_{df}^{ij} + 1920 R_{ae}^{bc} R_g^d R_{dh}^{ef} R_{ij}^{gh} R_{cf}^{ij} - 3840 R_{ae}^{bc} R_g^d R_{hi}^{ef} R_{cj}^{gh} R_{df}^{ij} - 3840 R_{ae}^{bc} R_i^d R_{cg}^{ef} R_{dj}^{gh} R_{fh}^{ij} +$$

$$+ 3840 R_{ae}^{bc} R_i^d R_{cg}^{ef} R_{fj}^{gh} R_{dh}^{ij} + 3840 R_{ae}^{bc} R_i^d R_{dg}^{ef} R_{cj}^{gh} R_{fh}^{ij} - 3840 R_{ae}^{bc} R_i^d R_{dg}^{ef} R_{fj}^{gh} R_{ch}^{ij} - 1920 R_{ae}^{bc} R_i^d R_{gh}^{ef} R_{cj}^{gh} R_{df}^{ij} +$$

$$+ 1920 R_{ae}^{bc} R_i^d R_{gh}^{ef} R_{dj}^{gh} R_{cf}^{ij} - 1920 R_{ae}^{bc} R_i^d R_{gh}^{ef} R_{fj}^{gh} R_{cd}^{ij} + 3840 R_{ae}^{bc} R_i^d R_{gj}^{ef} R_{cd}^{gh} R_{fh}^{ij} - 3840 R_{ae}^{bc} R_i^d R_{gj}^{ef} R_{cf}^{gh} R_{dh}^{ij} -$$

$$- 1920 R_{ae}^{cd} R_c^b R_{dg}^{ef} R_{ij}^{gh} R_{fh}^{ij} + 960 R_{ae}^{cd} R_c^b R_{gh}^{ef} R_{ij}^{gh} R_{df}^{ij} - 3840 R_{ae}^{cd} R_c^b R_{gi}^{ef} R_{dj}^{gh} R_{fh}^{ij} + 240 R_{ae}^{cd} R_f^b R_{cd}^{ef} R_{ij}^{gh} R_{gh}^{ij} -$$

$$- 1920 R_{ae}^{cd} R_f^b R_{cg}^{ef} R_{ij}^{gh} R_{dh}^{ij} + 480 R_{ae}^{cd} R_f^b R_{gh}^{ef} R_{ij}^{gh} R_{cd}^{ij} - 1920 R_{ae}^{cd} R_f^b R_{gi}^{ef} R_{cj}^{gh} R_{dh}^{ij} + 960 R_{ag}^{cd} R_e^b R_{cd}^{ef} R_{ij}^{gh} R_{fh}^{ij} +$$

$$+ 1920 R_{ag}^{cd} R_e^b R_{ch}^{ef} R_{ij}^{gh} R_{df}^{ij} + 1920 R_{ag}^{cd} R_e^b R_{hi}^{ef} R_{fj}^{gh} R_{cd}^{ij} + 3840 R_{ai}^{cd} R_e^b R_{cg}^{ef} R_{dj}^{gh} R_{fh}^{ij} - 3840 R_{ai}^{cd} R_e^b R_{cg}^{ef} R_{fj}^{gh} R_{dh}^{ij} +$$

$$+ 1920 R_{ai}^{cd} R_e^b R_{gh}^{ef} R_{cj}^{gh} R_{df}^{ij} + 960 R_{ai}^{cd} R_e^b R_{gh}^{ef} R_{fj}^{gh} R_{cd}^{ij} + 3840 R_{ai}^{cd} R_e^b R_{gj}^{ef} R_{cf}^{gh} R_{dh}^{ij} + 1920 R_{ae}^{cd} R_c^e R_{dg}^{bf} R_{ij}^{gh} R_{fh}^{ij} -$$

$$- 960 R_{ae}^{cd} R_c^e R_{gh}^{bf} R_{ij}^{gh} R_{df}^{ij} - 3840 R_{ae}^{cd} R_c^e R_{gi}^{bf} R_{fj}^{gh} R_{dh}^{ij} - 240 R_{ae}^{cd} R_f^e R_{cd}^{bf} R_{ij}^{gh} R_{gh}^{ij} + 1920 R_{ae}^{cd} R_f^e R_{cg}^{bf} R_{ij}^{gh} R_{dh}^{ij} -$$

$$- 480 R_{ae}^{cd} R_f^e R_{gh}^{bf} R_{ij}^{gh} R_{cd}^{ij} + 1920 R_{ae}^{cd} R_f^e R_{gi}^{bf} R_{cj}^{gh} R_{dh}^{ij} + 960 R_{ae}^{cd} R_g^e R_{cd}^{bf} R_{ij}^{gh} R_{fh}^{ij} + 1920 R_{ae}^{cd} R_g^e R_{ch}^{bf} R_{ij}^{gh} R_{df}^{ij} +$$

$$+ 1920 R_{ae}^{cd} R_g^e R_{hi}^{bf} R_{fj}^{gh} R_{cd}^{ij} + 3840 R_{ae}^{cd} R_i^e R_{cg}^{bf} R_{dj}^{gh} R_{fh}^{ij} - 3840 R_{ae}^{cd} R_i^e R_{cg}^{bf} R_{fj}^{gh} R_{dh}^{ij} + 1920 R_{ae}^{cd} R_i^e R_{gh}^{bf} R_{cj}^{gh} R_{df}^{ij} +$$

$$+ 960 R_{ae}^{cd} R_i^e R_{gh}^{bf} R_{fj}^{gh} R_{cd}^{ij} + 3840 R_{ae}^{cd} R_i^e R_{gj}^{bf} R_{cf}^{gh} R_{dh}^{ij} + 480 R_{af}^{cd} R_c^e R_{de}^{bf} R_{ij}^{gh} R_{gh}^{ij} - 1920 R_{af}^{cd} R_c^e R_{dg}^{bf} R_{ij}^{gh} R_{eh}^{ij} +$$

$$+ 1920 R_{af}^{cd} R_c^e R_{eg}^{bf} R_{ij}^{gh} R_{dh}^{ij} + 960 R_{af}^{cd} R_c^e R_{gh}^{bf} R_{ij}^{gh} R_{de}^{ij} - 3840 R_{af}^{cd} R_c^e R_{gi}^{bf} R_{dj}^{gh} R_{eh}^{ij} - 960 R_{af}^{cd} R_g^e R_{cd}^{bf} R_{ij}^{gh} R_{eh}^{ij} +$$

$$+ 1920 R_{af}^{cd} R_g^e R_{ce}^{bf} R_{ij}^{gh} R_{dh}^{ij} - 1920 R_{af}^{cd} R_g^e R_{ch}^{bf} R_{ij}^{gh} R_{de}^{ij} - 960 R_{af}^{cd} R_g^e R_{eh}^{bf} R_{ij}^{gh} R_{cd}^{ij} - 3840 R_{af}^{cd} R_g^e R_{hi}^{bf} R_{cj}^{gh} R_{de}^{ij} -$$

$$- 1920 R_{af}^{cd} R_g^e R_{hi}^{bf} R_{ej}^{gh} R_{cd}^{ij} - 3840 R_{af}^{cd} R_i^e R_{cg}^{bf} R_{dj}^{gh} R_{eh}^{ij} + 3840 R_{af}^{cd} R_i^e R_{cg}^{bf} R_{ej}^{gh} R_{dh}^{ij} - 3840 R_{af}^{cd} R_i^e R_{eg}^{bf} R_{cj}^{gh} R_{dh}^{ij} -$$

$$- 1920 R_{af}^{cd} R_i^e R_{gh}^{bf} R_{cj}^{gh} R_{de}^{ij} - 960 R_{af}^{cd} R_i^e R_{gh}^{bf} R_{ej}^{gh} R_{cd}^{ij} - 1920 R_{ag}^{cd} R_c^e R_{de}^{bf} R_{ij}^{gh} R_{fh}^{ij} - 1920 R_{ag}^{cd} R_c^e R_{dh}^{bf} R_{ij}^{gh} R_{ef}^{ij} +$$

$$+ 1920 R_{ag}^{cd} R_c^e R_{eh}^{bf} R_{ij}^{gh} R_{df}^{ij} - 3840 R_{ag}^{cd} R_c^e R_{hi}^{bf} R_{fj}^{gh} R_{de}^{ij} + 960 R_{ag}^{cd} R_f^e R_{cd}^{bf} R_{ij}^{gh} R_{eh}^{ij} - 1920 R_{ag}^{cd} R_f^e R_{ce}^{bf} R_{ij}^{gh} R_{dh}^{ij} +$$

$$+ 1920 R_{ag}^{cd} R_f^e R_{ch}^{bf} R_{ij}^{gh} R_{de}^{ij} + 960 R_{ag}^{cd} R_f^e R_{eh}^{bf} R_{ij}^{gh} R_{cd}^{ij} + 3840 R_{ag}^{cd} R_f^e R_{hi}^{bf} R_{cj}^{gh} R_{de}^{ij} + 1920 R_{ag}^{cd} R_f^e R_{hi}^{bf} R_{ej}^{gh} R_{cd}^{ij} -$$

$$- 960 R_{ag}^{cd} R_h^e R_{cd}^{bf} R_{ij}^{gh} R_{ef}^{ij} + 1920 R_{ag}^{cd} R_h^e R_{ce}^{bf} R_{ij}^{gh} R_{df}^{ij} + 1920 R_{ag}^{cd} R_i^e R_{cd}^{bf} R_{ej}^{gh} R_{fh}^{ij} - 1920 R_{ag}^{cd} R_i^e R_{cd}^{bf} R_{fj}^{gh} R_{eh}^{ij} -$$

$$- 3840 R_{ag}^{cd} R_i^e R_{ce}^{bf} R_{dj}^{gh} R_{fh}^{ij} + 3840 R_{ag}^{cd} R_i^e R_{ce}^{bf} R_{fj}^{gh} R_{dh}^{ij} - 3840 R_{ag}^{cd} R_i^e R_{ch}^{bf} R_{dj}^{gh} R_{ef}^{ij} + 3840 R_{ag}^{cd} R_i^e R_{ch}^{bf} R_{ej}^{gh} R_{df}^{ij} -$$

$$- 3840 R_{ag}^{cd} R_i^e R_{ch}^{bf} R_{fj}^{gh} R_{de}^{ij} - 3840 R_{ag}^{cd} R_i^e R_{eh}^{bf} R_{cj}^{gh} R_{df}^{ij} - 1920 R_{ag}^{cd} R_i^e R_{eh}^{bf} R_{fj}^{gh} R_{cd}^{ij} + 3840 R_{ag}^{cd} R_i^e R_{hj}^{bf} R_{cf}^{gh} R_{de}^{ij} +$$

$$+ 1920 R_{ag}^{cd} R_i^e R_{hj}^{bf} R_{ef}^{gh} R_{cd}^{ij} - 3840 R_{ai}^{cd} R_c^e R_{dg}^{bf} R_{ej}^{gh} R_{fh}^{ij} + 3840 R_{ai}^{cd} R_c^e R_{dg}^{bf} R_{fj}^{gh} R_{eh}^{ij} + 3840 R_{ai}^{cd} R_c^e R_{eg}^{bf} R_{dj}^{gh} R_{fh}^{ij} -$$

$$- 3840 R_{ai}^{cd} R_c^e R_{eg}^{bf} R_{fj}^{gh} R_{dh}^{ij} - 1920 R_{ai}^{cd} R_c^e R_{gh}^{bf} R_{dj}^{gh} R_{ef}^{ij} + 1920 R_{ai}^{cd} R_c^e R_{gh}^{bf} R_{ej}^{gh} R_{df}^{ij} - 1920 R_{ai}^{cd} R_c^e R_{gh}^{bf} R_{fj}^{gh} R_{de}^{ij} -$$

$$- 3840 R_{ai}^{cd} R_c^e R_{gj}^{bf} R_{df}^{gh} R_{eh}^{ij} + 3840 R_{ai}^{cd} R_c^e R_{gj}^{bf} R_{ef}^{gh} R_{dh}^{ij} + 3840 R_{ai}^{cd} R_f^e R_{cg}^{bf} R_{dj}^{gh} R_{eh}^{ij} - 3840 R_{ai}^{cd} R_f^e R_{cg}^{bf} R_{ej}^{gh} R_{dh}^{ij} +$$

$$+ 3840 R_{ai}^{cd} R_f^e R_{eg}^{bf} R_{cj}^{gh} R_{dh}^{ij} + 1920 R_{ai}^{cd} R_f^e R_{gh}^{bf} R_{cj}^{gh} R_{de}^{ij} + 960 R_{ai}^{cd} R_f^e R_{gh}^{bf} R_{ej}^{gh} R_{cd}^{ij} + 3840 R_{ai}^{cd} R_g^e R_{ch}^{bf} R_{dj}^{gh} R_{ef}^{ij} -$$

$$- 3840 R_{ai}^{cd} R_g^e R_{ch}^{bf} R_{ej}^{gh} R_{df}^{ij} + 3840 R_{ai}^{cd} R_g^e R_{ch}^{bf} R_{fj}^{gh} R_{de}^{ij} + 3840 R_{ai}^{cd} R_g^e R_{eh}^{bf} R_{cj}^{gh} R_{df}^{ij} + 1920 R_{ai}^{cd} R_g^e R_{eh}^{bf} R_{fj}^{gh} R_{cd}^{ij} -$$

$$- 3840 R_{ai}^{cd} R_g^e R_{hj}^{bf} R_{cf}^{gh} R_{de}^{ij} - 1920 R_{ai}^{cd} R_g^e R_{hj}^{bf} R_{ef}^{gh} R_{cd}^{ij} + 3840 R_{ai}^{cd} R_j^e R_{cg}^{bf} R_{de}^{gh} R_{fh}^{ij} - 3840 R_{ai}^{cd} R_j^e R_{cg}^{bf} R_{df}^{gh} R_{eh}^{ij} +$$

$$+ 3840 R_{ai}^{cd} R_j^e R_{cg}^{bf} R_{ef}^{gh} R_{dh}^{ij} + 1920 R_{ai}^{cd} R_j^e R_{eg}^{bf} R_{cd}^{gh} R_{fh}^{ij} - 3840 R_{ai}^{cd} R_j^e R_{eg}^{bf} R_{cf}^{gh} R_{dh}^{ij} - 960 R_{ai}^{cd} R_j^e R_{gh}^{bf} R_{cd}^{gh} R_{ef}^{ij} +$$

$$+ 1920 R_{ai}^{cd} R_j^e R_{gh}^{bf} R_{ce}^{gh} R_{df}^{ij} - 1920 R_{ai}^{cd} R_j^e R_{gh}^{bf} R_{cf}^{gh} R_{de}^{ij} - 960 R_{ai}^{cd} R_j^e R_{gh}^{bf} R_{ef}^{gh} R_{cd}^{ij} + 1920 R_{ad}^{bc} R_{cf}^{de} R_{eh}^{fg} R_{jk}^{hi} R_{gi}^{jk} -$$

$$- 960\, R_{ad}^{bc} R_{cf}^{de} R_{hi}^{fg} R_{jk}^{hi} R_{eg}^{jk} + 3840\, R_{ad}^{bc} R_{cf}^{de} R_{hj}^{fg} R_{ek}^{hi} R_{gi}^{jk} + 240\, R_{ad}^{bc} R_{fg}^{de} R_{ce}^{fg} R_{jk}^{hi} R_{hi}^{jk} - 960\, R_{ad}^{bc} R_{fg}^{de} R_{ch}^{fg} R_{jk}^{hi} R_{ei}^{jk} +$$
$$+ 960\, R_{ad}^{bc} R_{fg}^{de} R_{eh}^{fg} R_{jk}^{hi} R_{ci}^{jk} + 480\, R_{ad}^{bc} R_{fg}^{de} R_{hi}^{fg} R_{jk}^{hi} R_{ce}^{jk} - 1920\, R_{ad}^{bc} R_{fg}^{de} R_{hj}^{fg} R_{ck}^{hi} R_{ei}^{jk} - 1920\, R_{ad}^{bc} R_{fh}^{de} R_{ce}^{fg} R_{jk}^{hi} R_{gi}^{jk} -$$
$$- 1920\, R_{ad}^{bc} R_{fh}^{de} R_{ci}^{fg} R_{jk}^{hi} R_{eg}^{jk} + 3840\, R_{ad}^{bc} R_{fh}^{de} R_{cj}^{fg} R_{ek}^{hi} R_{gi}^{jk} - 3840\, R_{ad}^{bc} R_{fh}^{de} R_{cj}^{fg} R_{gk}^{hi} R_{ei}^{jk} + 1920\, R_{ad}^{bc} R_{fh}^{de} R_{ei}^{fg} R_{jk}^{hi} R_{cg}^{jk} +$$
$$+ 3840\, R_{ad}^{bc} R_{fh}^{de} R_{ej}^{fg} R_{gk}^{hi} R_{ci}^{jk} - 1920\, R_{ad}^{bc} R_{fh}^{de} R_{ij}^{fg} R_{gk}^{hi} R_{ce}^{jk} - 80\, R_{ae}^{cd} R_{cd}^{be} R_{hi}^{fg} R_{jk}^{hi} R_{fg}^{jk} + 320\, R_{ae}^{cd} R_{cd}^{be} R_{hj}^{fg} R_{fk}^{hi} R_{gi}^{jk} -$$
$$- 1920\, R_{ae}^{cd} R_{cf}^{be} R_{dh}^{fg} R_{jk}^{hi} R_{gi}^{jk} + 960\, R_{ae}^{cd} R_{cf}^{be} R_{hi}^{fg} R_{jk}^{hi} R_{dg}^{jk} - 3840\, R_{ae}^{cd} R_{cf}^{be} R_{hj}^{fg} R_{dk}^{hi} R_{gi}^{jk} - 120\, R_{ae}^{cd} R_{fg}^{be} R_{cd}^{fg} R_{jk}^{hi} R_{hi}^{jk} +$$
$$+ 960\, R_{ae}^{cd} R_{fg}^{be} R_{ch}^{fg} R_{jk}^{hi} R_{di}^{jk} - 240\, R_{ae}^{cd} R_{fg}^{be} R_{hi}^{fg} R_{jk}^{hi} R_{cd}^{jk} + 960\, R_{ae}^{cd} R_{fg}^{be} R_{hj}^{fg} R_{ck}^{hi} R_{di}^{jk} + 960\, R_{ae}^{cd} R_{fh}^{be} R_{cd}^{fg} R_{jk}^{hi} R_{gi}^{jk} +$$
$$+ 1920\, R_{ae}^{cd} R_{fh}^{be} R_{ci}^{fg} R_{jk}^{hi} R_{dg}^{jk} - 1920\, R_{ae}^{cd} R_{fh}^{be} R_{cj}^{fg} R_{dk}^{hi} R_{gi}^{jk} + 3840\, R_{ae}^{cd} R_{fh}^{be} R_{ij}^{fg} R_{ck}^{hi} R_{dg}^{jk} + 960\, R_{ae}^{cd} R_{fh}^{be} R_{ij}^{fg} R_{gk}^{hi} R_{cd}^{jk} -$$
$$- 960\, R_{af}^{cd} R_{cd}^{be} R_{eh}^{fg} R_{jk}^{hi} R_{gi}^{jk} + 480\, R_{af}^{cd} R_{cd}^{be} R_{hi}^{fg} R_{jk}^{hi} R_{eg}^{jk} - 1920\, R_{af}^{cd} R_{cd}^{be} R_{hj}^{fg} R_{ek}^{hi} R_{gi}^{jk} + 480\, R_{af}^{cd} R_{cg}^{be} R_{de}^{fg} R_{jk}^{hi} R_{hi}^{jk} -$$
$$- 1920\, R_{af}^{cd} R_{cg}^{be} R_{dh}^{fg} R_{jk}^{hi} R_{ei}^{jk} + 1920\, R_{af}^{cd} R_{cg}^{be} R_{eh}^{fg} R_{jk}^{hi} R_{di}^{jk} + 960\, R_{af}^{cd} R_{cg}^{be} R_{hi}^{fg} R_{jk}^{hi} R_{de}^{jk} + 3840\, R_{af}^{cd} R_{cg}^{be} R_{hj}^{fg} R_{ek}^{hi} R_{di}^{jk} +$$
$$+ 960\, R_{af}^{cd} R_{gh}^{be} R_{cd}^{fg} R_{jk}^{hi} R_{ei}^{jk} - 1920\, R_{af}^{cd} R_{gh}^{be} R_{ce}^{fg} R_{jk}^{hi} R_{di}^{jk} + 1920\, R_{af}^{cd} R_{gh}^{be} R_{ci}^{fg} R_{jk}^{hi} R_{de}^{jk} - 3840\, R_{af}^{cd} R_{gh}^{be} R_{cj}^{fg} R_{dk}^{hi} R_{ei}^{jk} +$$
$$+ 960\, R_{af}^{cd} R_{gh}^{be} R_{ei}^{fg} R_{jk}^{hi} R_{cd}^{jk} - 3840\, R_{af}^{cd} R_{gh}^{be} R_{ej}^{fg} R_{ck}^{hi} R_{di}^{jk} + 3840\, R_{af}^{cd} R_{gh}^{be} R_{ij}^{fg} R_{ck}^{hi} R_{de}^{jk} + 1920\, R_{ah}^{cd} R_{cf}^{be} R_{de}^{fg} R_{jk}^{hi} R_{gi}^{jk} +$$
$$+ 1920\, R_{ah}^{cd} R_{cf}^{be} R_{di}^{fg} R_{jk}^{hi} R_{eg}^{jk} - 3840\, R_{ah}^{cd} R_{cf}^{be} R_{dj}^{fg} R_{ek}^{hi} R_{gi}^{jk} + 3840\, R_{ah}^{cd} R_{cf}^{be} R_{dj}^{fg} R_{gk}^{hi} R_{ei}^{jk} - 1920\, R_{ah}^{cd} R_{cf}^{be} R_{ei}^{fg} R_{jk}^{hi} R_{dg}^{jk} +$$
$$+ 3840\, R_{ah}^{cd} R_{cf}^{be} R_{ej}^{fg} R_{dk}^{hi} R_{gi}^{jk} - 3840\, R_{ah}^{cd} R_{cf}^{be} R_{ej}^{fg} R_{gk}^{hi} R_{di}^{jk} + 3840\, R_{ah}^{cd} R_{cf}^{be} R_{ij}^{fg} R_{dk}^{hi} R_{eg}^{jk} - 3840\, R_{ah}^{cd} R_{cf}^{be} R_{ij}^{fg} R_{ek}^{hi} R_{dg}^{jk} +$$
$$+ 3840\, R_{ah}^{cd} R_{cf}^{be} R_{ij}^{fg} R_{gk}^{hi} R_{de}^{jk} + 1920\, R_{ah}^{cd} R_{cf}^{be} R_{jk}^{fg} R_{de}^{hi} R_{gi}^{jk} - 1920\, R_{ah}^{cd} R_{cf}^{be} R_{jk}^{fg} R_{dg}^{hi} R_{ei}^{jk} + 1920\, R_{ah}^{cd} R_{cf}^{be} R_{jk}^{fg} R_{eg}^{hi} R_{di}^{jk} +$$
$$+ 480\, R_{ah}^{cd} R_{fg}^{be} R_{cd}^{fg} R_{jk}^{hi} R_{ei}^{jk} - 960\, R_{ah}^{cd} R_{fg}^{be} R_{ce}^{fg} R_{jk}^{hi} R_{di}^{jk} + 960\, R_{ah}^{cd} R_{fg}^{be} R_{ci}^{fg} R_{jk}^{hi} R_{de}^{jk} + 480\, R_{ah}^{cd} R_{fg}^{be} R_{ei}^{fg} R_{jk}^{hi} R_{cd}^{jk} +$$
$$+ 960\, R_{ah}^{cd} R_{fg}^{be} R_{ij}^{fg} R_{ek}^{hi} R_{cd}^{jk} - 960\, R_{ah}^{cd} R_{fi}^{be} R_{cd}^{fg} R_{jk}^{hi} R_{eg}^{jk} + 1920\, R_{ah}^{cd} R_{fi}^{be} R_{ce}^{fg} R_{jk}^{hi} R_{dg}^{jk} + 3840\, R_{ah}^{cd} R_{fi}^{be} R_{ej}^{fg} R_{ck}^{hi} R_{dg}^{jk} +$$
$$+ 1920\, R_{ah}^{cd} R_{fi}^{be} R_{ej}^{fg} R_{gk}^{hi} R_{cd}^{jk} + 1920\, R_{aj}^{cd} R_{fg}^{be} R_{ch}^{fg} R_{dk}^{hi} R_{ei}^{jk} - 1920\, R_{aj}^{cd} R_{fg}^{be} R_{ch}^{fg} R_{ek}^{hi} R_{di}^{jk} + 1920\, R_{aj}^{cd} R_{fg}^{be} R_{eh}^{fg} R_{ck}^{hi} R_{di}^{jk} +$$
$$+ 960\, R_{aj}^{cd} R_{fg}^{be} R_{hi}^{fg} R_{ck}^{hi} R_{de}^{jk} + 480\, R_{aj}^{cd} R_{fg}^{be} R_{hi}^{fg} R_{ek}^{hi} R_{cd}^{jk} + 1920\, R_{aj}^{cd} R_{fg}^{be} R_{hk}^{fg} R_{ce}^{hi} R_{di}^{jk} + 3840\, R_{aj}^{cd} R_{fh}^{be} R_{ce}^{fg} R_{dk}^{hi} R_{gi}^{jk} -$$
$$- 3840\, R_{aj}^{cd} R_{fh}^{be} R_{ce}^{fg} R_{gk}^{hi} R_{di}^{jk} + 3840\, R_{aj}^{cd} R_{fh}^{be} R_{ci}^{fg} R_{dk}^{hi} R_{eg}^{jk} + 3840\, R_{aj}^{cd} R_{fh}^{be} R_{ci}^{fg} R_{gk}^{hi} R_{de}^{jk} + 3840\, R_{aj}^{cd} R_{fh}^{be} R_{ei}^{fg} R_{ck}^{hi} R_{dg}^{jk} +$$
$$+ 1920\, R_{aj}^{cd} R_{fh}^{be} R_{ei}^{fg} R_{gk}^{hi} R_{cd}^{jk} - 1920\, R_{aj}^{cd} R_{fh}^{be} R_{ek}^{fg} R_{cd}^{hi} R_{gi}^{jk} + 3840\, R_{aj}^{cd} R_{fh}^{be} R_{ek}^{fg} R_{cg}^{hi} R_{di}^{jk} - 1920\, R_{aj}^{cd} R_{fh}^{be} R_{ik}^{fg} R_{cd}^{hi} R_{eg}^{jk}).$$

For a check, note that (1) $G_{(5)a}{}^{a} = \frac{n-10}{10} L_{(5)}$ and (2) the magnitudes of the numerical coefficients of $G_{(5)a}{}^{b}$ add up to $\frac{11!}{2^6 \times 5} = \frac{1}{10} \times 1{,}247{,}400 = 124{,}740$.

## APPENDIX. EXPRESSION FOR $L_{(5)}$

"In a certain sense $L_{(5)}$ forms a boundary, since the higher terms do not appear in the 'physically interesting' (supergravity, superstrings) dimensions < 12."
— Müller-Hoissen, F., [personal communication], tr. by author.

The coefficient $L_{(5)}$ of the quintic Lovelock Lagrangian is given by the formula

$$L_{(5)} = \frac{10!}{2^5} R_{[i_1 i_2}{}^{i_1 i_2} R_{i_3 i_4}{}^{i_3 i_4} R_{i_5 i_6}{}^{i_5 i_6} R_{i_7 i_8}{}^{i_7 i_8} R_{i_9 i_{10}]}{}^{i_9 i_{10}}, \tag{6}$$

which comprises 3,628,800 unique covariant index permutations, of which but 85—together with numerical coefficients—suffice for rendering a general expression for $L_{(5)}$, the final result (after substituting contractions and re-labeling indices) being given by

$$L_{(5)} = - R^5 + 40\, R^3 R_b^a R_a^b - 10\, R^3 R_{cd}^{ab} R_{ab}^{cd} - 160\, R^2 R_b^a R_c^b R_a^c + 240\, R^2 R_c^a R_d^b R_{ab}^{cd} + 240\, R^2 R_b^a R_{de}^{bc} R_{ac}^{de} + 20\, R^2 R_{cd}^{ab} R_{ef}^{cd} R_{ab}^{ef} - \tag{7}$$
$$- 80\, R^2 R_{ce}^{ab} R_{af}^{cd} R_{bd}^{ef} - 240\, R R_b^a R_a^b R_d^c R_c^d + 480\, R R_b^a R_c^b R_d^c R_a^d - 1920\, R R_b^a R_d^b R_e^c R_{ac}^{de} + 120\, R R_b^a R_a^b R_{ef}^{cd} R_{cd}^{ef} -$$
$$- 960\, R R_b^a R_c^b R_{ef}^{cd} R_{ad}^{ef} - 480\, R R_c^a R_d^b R_{ef}^{cd} R_{ab}^{ef} + 960\, R R_c^a R_e^b R_{af}^{cd} R_{bd}^{ef} - 960\, R R_c^a R_e^b R_{bf}^{cd} R_{ad}^{ef} + 960\, R R_b^a R_{ad}^{bc} R_{fg}^{de} R_{ce}^{fg} -$$
$$- 480\, R R_b^a R_{de}^{bc} R_{fg}^{de} R_{ac}^{fg} + 1920\, R R_b^a R_{df}^{bc} R_{ag}^{de} R_{ce}^{fg} - 15\, R R_{cd}^{ab} R_{ab}^{cd} R_{gh}^{ef} R_{ef}^{gh} + 240\, R R_{cd}^{ab} R_{ae}^{cd} R_{gh}^{ef} R_{bf}^{gh} -$$
$$- 30\, R R_{cd}^{ab} R_{ef}^{cd} R_{gh}^{ef} R_{ab}^{gh} + 480\, R R_{cd}^{ab} R_{eg}^{cd} R_{ah}^{ef} R_{bf}^{gh} - 240\, R R_{ce}^{ab} R_{ag}^{cd} R_{bh}^{ef} R_{df}^{gh} + 480\, R R_{ce}^{ab} R_{ag}^{cd} R_{dh}^{ef} R_{bf}^{gh} +$$
$$+ 640\, R_b^a R_a^b R_d^c R_e^d R_c^e - 768\, R_b^a R_c^b R_d^c R_e^d R_a^e - 960\, R_b^a R_a^b R_e^c R_f^d R_{cd}^{ef} + 3840\, R_b^a R_c^b R_e^c R_f^d R_{ad}^{ef} + 1920\, R_b^a R_e^b R_d^c R_f^d R_{ac}^{ef} -$$
$$- 960\, R_b^a R_a^b R_d^c R_{fg}^{de} R_{ce}^{fg} - 160\, R_b^a R_c^b R_a^c R_{fg}^{de} R_{de}^{fg} + 1920\, R_b^a R_c^b R_d^c R_{fg}^{de} R_{ae}^{fg} + 1920\, R_b^a R_d^b R_e^c R_{fg}^{de} R_{ac}^{fg} -$$
$$- 3840\, R_b^a R_d^b R_f^c R_{ag}^{de} R_{ce}^{fg} + 3840\, R_b^a R_d^b R_f^c R_{cg}^{de} R_{ae}^{fg} + 1920\, R_d^a R_f^b R_e^c R_{bg}^{de} R_{ac}^{fg} + 3840\, R_d^a R_f^b R_g^c R_{ab}^{de} R_{ce}^{fg} -$$
$$- 80\, R_b^a R_a^b R_{ef}^{cd} R_{gh}^{ef} R_{cd}^{gh} + 320\, R_b^a R_a^b R_{eg}^{cd} R_{ch}^{ef} R_{df}^{gh} - 1920\, R_b^a R_c^b R_{ae}^{cd} R_{gh}^{ef} R_{df}^{gh} + 960\, R_b^a R_c^b R_{ef}^{cd} R_{gh}^{ef} R_{ad}^{gh} -$$
$$- 3840\, R_b^a R_c^b R_{eg}^{cd} R_{ah}^{ef} R_{df}^{gh} + 240\, R_c^a R_d^b R_{ab}^{cd} R_{gh}^{ef} R_{ef}^{gh} - 1920\, R_c^a R_d^b R_{ae}^{cd} R_{gh}^{ef} R_{bf}^{gh} + 480\, R_c^a R_d^b R_{ef}^{cd} R_{gh}^{ef} R_{ab}^{gh} -$$
$$- 1920\, R_c^a R_d^b R_{eg}^{cd} R_{ah}^{ef} R_{bf}^{gh} - 1920\, R_c^a R_e^b R_{ab}^{cd} R_{gh}^{ef} R_{df}^{gh} - 1920\, R_c^a R_e^b R_{af}^{cd} R_{gh}^{ef} R_{bd}^{gh} - 3840\, R_c^a R_g^b R_{ae}^{cd} R_{bh}^{ef} R_{df}^{gh} +$$
$$+ 1920\, R_c^a R_g^b R_{ae}^{cd} R_{dh}^{ef} R_{bf}^{gh} + 3840\, R_c^a R_g^b R_{be}^{cd} R_{ah}^{ef} R_{df}^{gh} - 1920\, R_c^a R_g^b R_{be}^{cd} R_{dh}^{ef} R_{af}^{gh} + 1920\, R_c^a R_g^b R_{ef}^{cd} R_{bh}^{ef} R_{ad}^{gh} +$$
$$+ 1920\, R_c^a R_g^b R_{eh}^{cd} R_{ab}^{ef} R_{df}^{gh} + 1920\, R_b^a R_{ad}^{bc} R_{cf}^{de} R_{hi}^{fg} R_{eg}^{hi} - 960\, R_b^a R_{ad}^{bc} R_{fg}^{de} R_{hi}^{fg} R_{ce}^{hi} + 3840\, R_b^a R_{ad}^{bc} R_{fh}^{de} R_{ci}^{fg} R_{eg}^{hi} +$$
$$+ 240\, R_b^a R_{de}^{bc} R_{ac}^{de} R_{hi}^{fg} R_{fg}^{hi} - 960\, R_b^a R_{de}^{bc} R_{af}^{de} R_{hi}^{fg} R_{cg}^{hi} + 960\, R_b^a R_{de}^{bc} R_{cf}^{de} R_{hi}^{fg} R_{ag}^{hi} + 480\, R_b^a R_{de}^{bc} R_{fg}^{de} R_{hi}^{fg} R_{ac}^{hi} -$$
$$- 1920\, R_b^a R_{de}^{bc} R_{fh}^{de} R_{ai}^{fg} R_{cg}^{hi} - 1920\, R_b^a R_{df}^{bc} R_{ac}^{de} R_{hi}^{fg} R_{eg}^{hi} - 1920\, R_b^a R_{df}^{bc} R_{ag}^{de} R_{hi}^{fg} R_{ce}^{hi} + 3840\, R_b^a R_{df}^{bc} R_{ah}^{de} R_{ci}^{fg} R_{eg}^{hi} -$$
$$- 3840\, R_b^a R_{df}^{bc} R_{ah}^{de} R_{ei}^{fg} R_{cg}^{hi} + 1920\, R_b^a R_{df}^{bc} R_{cg}^{de} R_{hi}^{fg} R_{ae}^{hi} + 3840\, R_b^a R_{df}^{bc} R_{ch}^{de} R_{ei}^{fg} R_{ag}^{hi} - 1920\, R_b^a R_{df}^{bc} R_{gh}^{de} R_{ei}^{fg} R_{ac}^{hi} +$$
$$+ 20\, R_{cd}^{ab} R_{ab}^{cd} R_{gh}^{ef} R_{ij}^{gh} R_{ef}^{ij} - 80\, R_{cd}^{ab} R_{ab}^{cd} R_{gi}^{ef} R_{ej}^{gh} R_{fh}^{ij} + 480\, R_{cd}^{ab} R_{ae}^{cd} R_{bg}^{ef} R_{ij}^{gh} R_{fh}^{ij} - 480\, R_{cd}^{ab} R_{ae}^{cd} R_{gh}^{ef} R_{ij}^{gh} R_{bf}^{ij} +$$
$$+ 1920\, R_{cd}^{ab} R_{ae}^{cd} R_{gi}^{ef} R_{bj}^{gh} R_{fh}^{ij} + 24\, R_{cd}^{ab} R_{ef}^{cd} R_{gh}^{ef} R_{ij}^{gh} R_{ab}^{ij} - 480\, R_{cd}^{ab} R_{ef}^{cd} R_{gi}^{ef} R_{aj}^{gh} R_{bh}^{ij} - 480\, R_{cd}^{ab} R_{eg}^{cd} R_{ah}^{ef} R_{ij}^{gh} R_{bf}^{ij} +$$
$$+ 960\, R_{cd}^{ab} R_{eg}^{cd} R_{ai}^{ef} R_{bj}^{gh} R_{fh}^{ij} - 1920\, R_{cd}^{ab} R_{eg}^{cd} R_{ai}^{ef} R_{fj}^{gh} R_{bh}^{ij} + 1920\, R_{ce}^{ab} R_{af}^{cd} R_{gi}^{ef} R_{bj}^{gh} R_{dh}^{ij} - 384\, R_{ce}^{ab} R_{ag}^{cd} R_{bi}^{ef} R_{dj}^{gh} R_{fh}^{ij} +$$
$$+ 1920\, R_{ce}^{ab} R_{ag}^{cd} R_{bi}^{ef} R_{fj}^{gh} R_{dh}^{ij} - 1920\, R_{ce}^{ab} R_{ag}^{cd} R_{di}^{ef} R_{fj}^{gh} R_{bh}^{ij} - 768\, R_{ce}^{ab} R_{fg}^{cd} R_{hi}^{ef} R_{aj}^{gh} R_{bd}^{ij}.$$

For a check, note that the magnitudes of the numerical coefficients of $L_{(5)}$ add up to $\frac{10!}{2^5} = 113{,}400$.